\documentclass[useAMS,usenatbib]{mn2e}
\usepackage{aas_macros,graphicx,times,multirow}

\newcommand{\xmm}{{\em XMM-Newton}}
\newcommand{\chan}{{\em Chandra}}
\def \psr{PSR\, J0007+7303}

\title[Optical observations of \psr]{Deep optical observations of the $\gamma$-ray pulsar \psr\ in the CTA 1 supernova remnant}
\author[R. P. Mignani, A. de Luca.,  N.Rea, et al. ]
{\parbox{\textwidth}{R. P. Mignani$^{1,2,3}$\thanks{E-mail: rm2@mssl.ucl.ac.uk}, 
A. de Luca$^{1,4}$,
N. Rea$^{5}$, 
A. Shearer$^{6}$, 
S. Collins$^{6}$, 
D. F. Torres$^{5}$, 
D. Hadasch$^{5}$,
A. Caliandro$^{5}$} \\ \\
$^{1}$ INAF - Istituto di Astrofisica Spaziale e Fisica Cosmica Milano, via E. Bassini 15, 20133, Milano, Italy\\
$^{2}$ Mullard Space Science Laboratory, University College London, Holmbury St. Mary, Dorking, Surrey, RH5 6NT, UK\\
$^{3}$ Kepler Institute of Astronomy, University of Zielona G\'ora, Lubuska 2, 65-265, Zielona G\'ora, Poland \\
$^{4}$ INFN - Istituto Nazionale di Fisica Nucleare, sezione di Pavia, via A. Bassi 6, 27100, Pavia, Italy \\
$^{5}$  Institut de Ci\'encies de lÕEspai (CSIC-IEEC), Campus UAB, Facultat de Ci\'encies, Torre C5-parell, E-08193 Barcelona, Spain\\
$^{6}$ Centre for Astronomy, National University of Ireland, Newcastle Road, Galway, Ireland
}
\begin{document}

\date{Accepted 1988 December 15. Received 1988 December 14; in original form 1988 October 11}

\pagerange{\pageref{firstpage}--\pageref{lastpage}} \pubyear{2002}

\maketitle

\label{firstpage}

\begin{abstract}
The {\em Fermi} Large Area Telescope (LAT) discovered the time signature of a radio-silent pulsar coincident with RX\, J0007.0+7302, a plerion-like X-ray source at the centre of the CTA\, 1 supernova remnant. The inferred timing parameters of the $\gamma$-ray pulsar \psr\ ($P$=315.8 ms; $\dot{P} \sim3.6\times10^{-13}$ s s$^{-1}$) point to a Vela-like neutron star, with an age comparable to that of CTA\, 1. The \psr\  low distance ($\sim 1.4$ kpc), interstellar absorption ($A_V\sim 1.6$), and relatively high energy loss rate ($\dot{E} \sim 4.5\times10^{35}$ erg s$^{-1}$), make it a suitable candidate for an optical follow-up. Here, we present  deep optical observations of \psr.    The pulsar is not detected in the Gran Telescopio Canarias (GTC)  images down to a limit of $r' \sim 27.6$ ($3\sigma$), the deepest ever obtained for this pulsar, while  William Herschel Telescope (WHT)  images yield a limit of $V\sim 26.9$. Our $r'$-band  limit corresponds to an optical emission efficiency $\eta_{opt} \equiv L_{opt}/\dot{E} \la 9.4 \times 10^{-8}$. This limit is more constraining than those derived for other Vela-like pulsars, but is still above the measured optical efficiency of the Vela pulsar. We compared the 
optical  upper limits with the extrapolation of the \xmm\ X-ray spectrum and found that the optical emission is compatible with the extrapolation of the X-ray power-law component, at variance with what is observed, e.g. in the Vela pulsar.
\end{abstract}


\section{Introduction}

The  Large Area Telescope (LAT) on {\em Fermi}, successfully launched on June 11, 2008 is revolutionising our view of the  $\gamma$-ray sky, thanks to its large  collecting area and outstanding performances at energies above 1 GeV.  During its activation phase, the {\em Fermi}/LAT discovered the timing signature of a radio-silent pulsar within CTA\, 1 (Abdo et al.\ 2008). This is  a $\sim 5$--15 kyr old supernova remnant  (SNR) located at $1.4 \pm 0.3$ kpc (Pineault et al.\ 1993), recently detected also at TeV energies (Aliu et al.\ 2012).  The $\gamma$-ray source, already detected by the Energetic Gamma Ray Experiment Telescope (EGRET) aboard the {\em Compton Gamma-ray Observatory} ({\em CGRO}) as 3EG\, J0010+7309,  
coincides with RX\,  J0007.0+7302, a composite, plerion-like X-ray source  studied by {\em ROSAT} (Seward et al.\ 1995), {\em ASCA} (Slane et al.\ 1997), {\em Chandra} (Halpern et al.\ 2004), and {\em XMM-Newton} (Slane et al.\ 2004), and for a long time suspected to be a neutron star. Deep radio observation assessed that a putative pulsar should have a luminosity lower than the faintest known radio pulsar (Halpern et al.\ 2004; Camilo et al.\ 2009).  A blind search on $\gamma$-ray photons collected by {\em Fermi}/LAT (Abdo et al.\ 2008) has unveiled a clear periodicity at $\sim315.8$ ms, with a  period derivative  $\dot{P} \sim3.6\times10^{-13}$ s s$^{-1}$, corresponding to  an overall energy loss $\dot{E} \sim 4.5\times10^{35}$ erg s$^{-1}$, a dipolar magnetic field $B \sim 1.08 \times 10^{13}$ G,  and a characteristic age of $\sim13.9$ kyr, consistent with that of CTA\, 1. 
A recent, very deep (120 ks), {\em XMM-Newton} observation of the pulsar in CTA\, 1 (Caraveo et al.\ 2010) allowed us to characterise the X-ray phenomenology (both temporal and spectral properties) of  this source in a definitive way. 

After the "first light" discovery of \psr, more than 100 $\gamma$-ray pulsars have been detected by the {\em Fermi}/LAT (Nolan et al.\ 2012). 
In the optical, follow-up observations of {\em Fermi} pulsars with either the {\em Hubble Space Telescope} ({\em HST}) or 8m-class telescopes have been performed in a few cases only (Mignani et al.\  2010a,b; 2011; 2012; Razzano et al.\ 2012; Shearer et al.\ 2012).   Owing to its energetics, proximity, low absorption ($N_{\rm H}=1.66^{+0.89}_{-0.76} ~10^{21}$cm$^{-2}$; Caraveo et al.\ 2010) and off-plane position ($b\sim10^{\circ}$), \psr\  is a good candidate for a detection at optical wavelengths, indeed one of the best within the new emerging class of $\gamma$-ray  bright, radio-silent pulsars.  No observations of the pulsar with 8m-class telescopes have been performed yet. Images collected with the 2.4 m Hiltner telescope at the MDM Observatory, prior to the discovery of the $\gamma$-ray pulsar,  set upper limits of B$\sim25.4$ and V$\sim24.9$ (Halpern et al.\ 2004) at the pulsar \chan\ position,
way too bright than the expected source magnitude.  

We used different telescopes at the Roque de Los Muchachos Observatory (La Palma, Spain), including the  10.4m Gran Telescopio Canarias (GTC), to collect deep images of the \psr\ field  and search for its optical counterpart. The paper is structured as follows: observations and data reduction are described in Sectn. 2, while the results are presented and discussed in Sectn. 3 and 4, respectively.

\section{Observations and data reduction}

We obtained deep observations of the \psr\ field with the GTC on  August 28 and 29, 2011 under   programme GTC21-11A\_002 (PI. N. Rea).  We observed \psr\ with the Optical System for Imaging and low Resolution Integrated Spectroscopy (OSIRIS). The instrument is equipped with a two-chip E2V CCD detector with a nominal field--of--view (FoV) of $7\farcm8\times8\farcm5$ that is actually decreased to $7\arcmin \times 7\arcmin$ due to the vignetting of Chip 1. The unbinned pixel size of the CCD is 0\farcs125. In total, we took three sequences of 15 exposures in the Sloan $r'$  band ($\lambda=6410$ \AA; $\Delta \lambda=1760$\AA) with exposure time of 140 s,  to minimise the saturation of bright stars in the field and remove cosmic ray hits. Exposures were dithered by 10\arcsec\ steps in right ascension and declination to correct for the fringing. The pulsar was positioned at the nominal aim point in Chip 2. Observations were performed in dark and clear sky conditions, with an average airmass of 1.47, due to the high declination of our target, and a seeing of $\sim 0\farcs9$.  

We also observed \psr\ with  the 2.5m Isaac Newton Telescope (INT), the 3.5m Telescopio Nazionale Galileo (TNG), and the 4m William Herschel Telescope (WHT) at the Roque de Los Muchachos Observatory between August 1 and October 13, 2009, under the International Time Proposal ITP02 (PI. A. Shearer) as part of a pilot survey of {\em Fermi} pulsar fields (Collins et al.\ 2011). INT observations were performed with the Wide Field Camera (WFC), a mosaic of four  EEV CCDs ($34\farcm2\times34\farcm2$ FoV; 0\farcs33/pixel). Exposures were taken in dark sky conditions (1.40 airmass; 1\farcs2 seeing) through the Harris $R$-band ($\lambda=6380$\AA; $\Delta \lambda=1520$\AA) for a total integration time of 1800 s.  TNG observations were performed  with the  DOLORES (Device Optimized for the LOw RESolution) camera,  a single-chip E2V CCD ($8\farcm6 \times 8\farcm6$ FoV; 0\farcs252/pixel). Exposures through both the standard Johnson $V$ and $R$ bands were taken in dark sky conditions (1.43 airmass; 1\farcs6 seeing) for a total integration time of 2760 and 9900 s, respectively.   Finally, WHT observations  were also performed in dark  sky conditions (1.45 airmass; 1\farcs2 seeing) with the Prime Focus Camera (PFC), a mosaic of two EEV CCDs ($16\farcm2\times16\farcm2$ FoV; 0\farcs24/pixel). Exposures were taken  through both the Harris $V$ ($\lambda=5437$\AA; $\Delta \lambda=957$\AA) and R  ($\lambda=6408$\AA; $\Delta \lambda=1562$\AA) bands for a total integration time of 3600s per band. The observation summary is given in Table \ref{obs}. In all runs, images of Landolt standard star fields were taken for photometric calibration, together with twilight sky flat field and day-time calibration frames.

   \begin{table}
\centering
\caption{Summary of the optical observations of \psr\ including: the observing date, band, integration time (T), airmass and seeing.}
\label{obs}
\begin{tabular}{lccccc} \hline
Telescope & Date      &  Band & T   & airmass & seeing\\
                  & yyyy-mm-dd  &    &  (s)  & & (\arcsec)\\ \hline
GTC           & 2011-08-28  &  $r'$ & 4200 & 1.47 & 0.8 \\ \
                 & 2011-08-29  &  $r'$ & 2100 & 1.46 & 1.2 \\ \hline
WHT             & 2009-10-13 &  $V$ & 3600 & 1.42 & 1.3\\ 
            & 2009-10-13 &  $R$ & 3600 & 1.49  & 1.0 \\               \hline
TNG              & 2009-08-18 &  $V$ & 2760 & 1.44 & 1.7\\ 
            & 2009-08-18 &  $R$ & 2700 & 1.44  & 1.7\\
                & 2009-08-20 &  $R$ & 7200 & 1.42 & 1.4\\ \hline
INT              & 2009-08-01 &  $R$ & 1800 & 1.40 & 1.2\\ \hline
\end{tabular}
\vspace{0.5cm}
\end{table}

 We reduced our data  (bias and dark subtraction, flat-field correction)  using standard tools in the {\sc IRAF} package {\sc ccdred}. Single dithered exposures were then aligned, stacked, and filtered by cosmic rays using the task {\tt drizzle}. 
 We computed the astrometry calibration using the {\em wcstools}\footnote{http://tdc-www.harvard.edu/wcstools/} suite of programs, matching the sky coordinates of stars selected from the Two Micron All Sky Survey (2MASS) All-Sky Catalog of Point Sources (Skrutskie et al.\ 2006)
 with their pixel coordinates computed by {\em Sextractor} (Bertin \&Arnouts 1996).  After iterating the matching process applying a $\sigma$-clipping selection to filter out obvious mismatches, high-proper motion stars, and false detections, we obtained mean residuals of $\sim 0\farcs2$ in the radial direction, using up to 30 bright, but non-saturated, 
 2MASS stars. To this value we added in quadrature the uncertainty $\sigma_{tr}$ = 0\farcs08 of the image registration  on the
 2MASS reference frame. This is given by $\sigma_{tr}$=$\sqrt{n/N_{S}}\sigma_{\rm S}$ (e.g., Lattanzi et al.\ 1997), where $N_{S}$ is the number of stars used to compute the astrometric solution, $n$=5 is the number of free parameters in the sky--to--image transformation model, $\sigma_{\rm S} \sim 0\farcs2$ is the mean absolute position error of  2MASS (Skrutskie et al.\ 2006) for stars in the magnitude range  $15.5 \le K \le 13$.
 After accounting for the 0\farcs015  uncertainty on the link of 
2MASS to the International Celestial Reference Frame  (Skrutskie et al.\ 2006),
we ended up with an overall accuracy of $\sim$0\farcs22 on our absolute astrometry.

\begin{figure}
\centering
{\includegraphics[height=8cm,clip=]{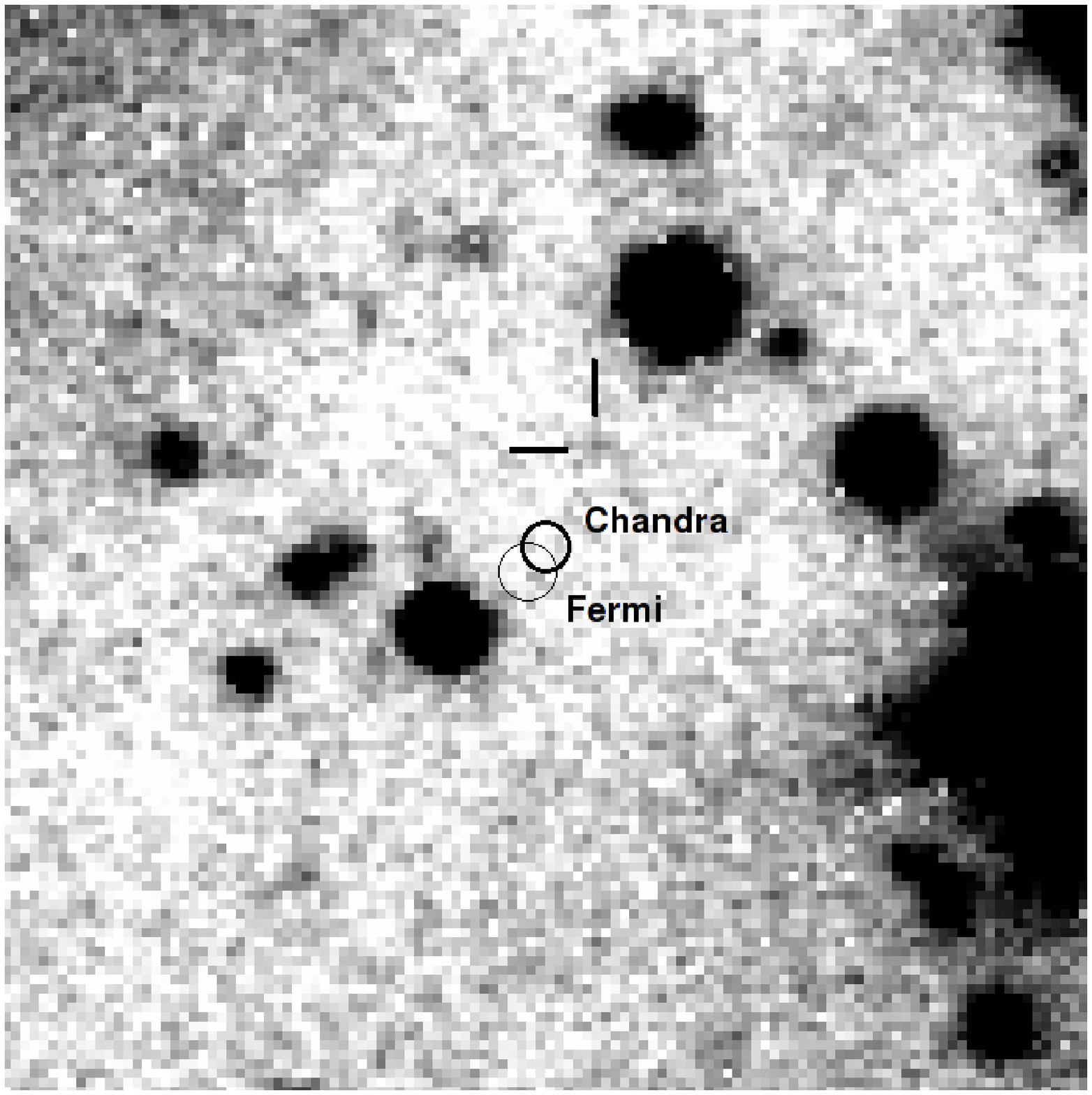}}
{\includegraphics[height=8cm, clip=]{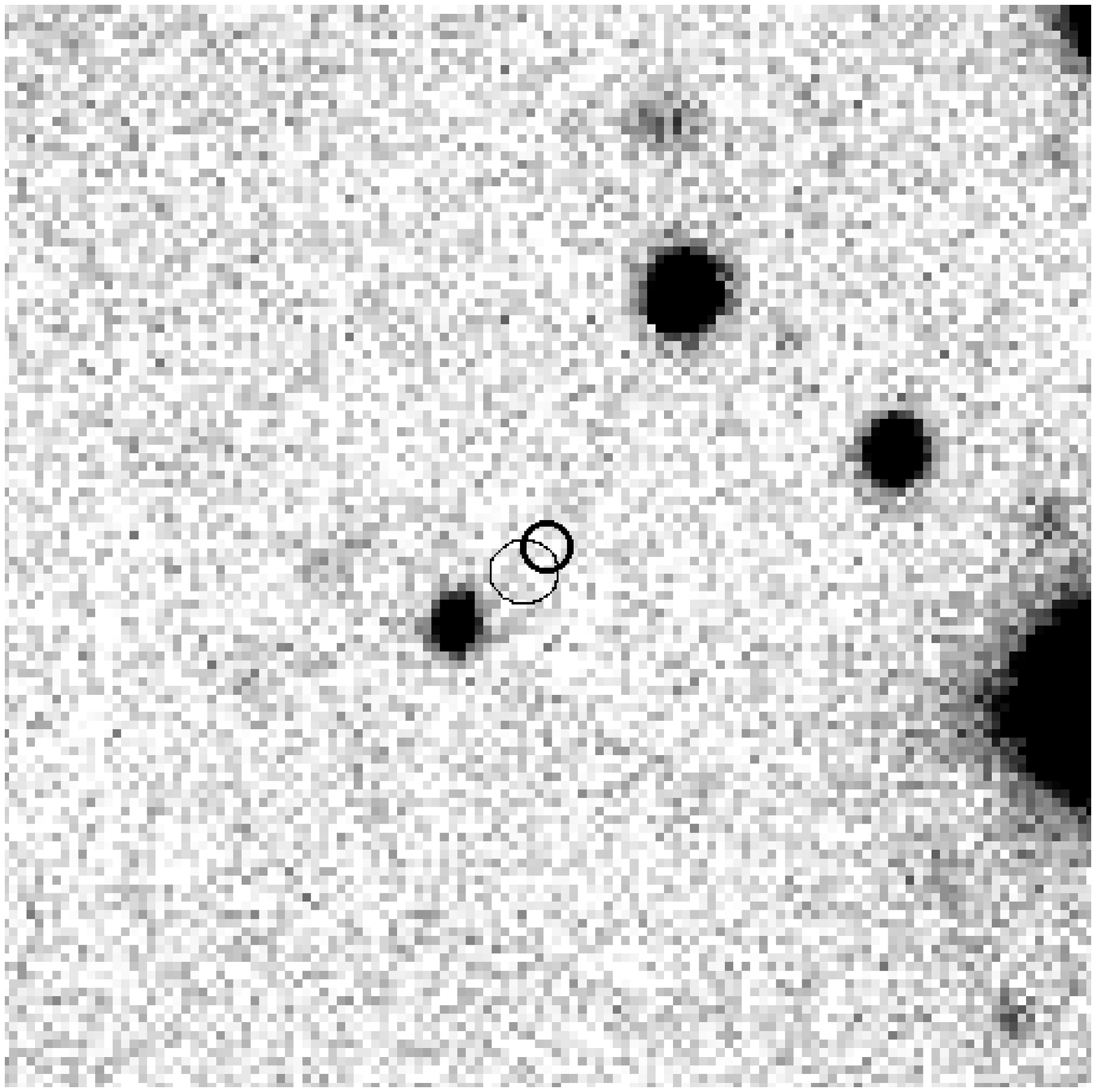}}
\caption{\label{fc} 
{\em Top panel:} $30\arcsec \times 30\arcsec$ GTC/OSIRIS image of the \psr\ field  ($r'$ band; 6300 s). North to the top, east to the left. The pulsar \chan\ and {\em Fermi}/LAT (Abdo et al.\ 2012) positions are marked by the two circles and labelled. Their size (0\farcs65 and 0\farcs83 radius, respectively) accounts for both the nominal coordinate uncertainties and the accuracy of the astrometry calibration of the optical images (see Sectn.\ 2). A faint ($R'\approx27$) object detected $\sim 3\arcsec$ north west of the \chan\ position is marked by the two ticks. {\em Bottom panel:} Same region observed in the V band (3600 s) with the WHT/PFC.}
\end{figure}

\section{Results}

A \chan\ X-ray position for \psr\ had been reported by Halpern et al.\ (2004), i.e. $\alpha =00^{\rm h}  07^{\rm m} 01\fs56$; $\delta  = +73^\circ 03\arcmin 08\farcs10$ (J2000). These coordinates had been bore-sight corrected using matches with the United States Naval Observatory (USNO) A2.0 catalog (Monet et al.\ 1998) and have a reported accuracy of 0\farcs1 with respect to the optical reference frame. We verified the position of Halpern et al.\ (2004) using the same \chan\ data set after pipeline re-processing with  the newest calibration data sets and using more recent optical/infrared catalogues to re-calibrate the astrometry.  We retrieved "level 1" data from the \chan\ X-ray Science Archive and reprocessed them with the \chan\ Interactive Analysis of Observations Software (CIAO v4.4)\footnote{{\texttt http://cxc.harvard.edu/ciao/index.html}} using the {\tt chandra\_repro} script\footnote{{\texttt http://cxc.harvard.edu/ciao/ahelp/chandra\_repro.html}}. The pulsar is imaged on the ACIS-I CCD array.  We extracted an image in the 0.3--8 keV energy range with the native ACIS detector resolution (0\farcs492/pixel) and ran a source detection using the {\tt wavdetect}\footnote{{\texttt http://cxc.harvard.edu/ciao/threads/wavdetect/}} task with wavelet scales ranging from 1 to 16 pixels with $\sqrt{2}$ steps. We cross-correlated the derived source list with stars in the 2MASS catalogue (Skrutskie et al.\ 2006)  and obtained 6 matches using a 1\arcsec\ cross-correlation radius.  
However, we could determine no statistically significant transformations to optimise the ACIS astrometric solution,  with best-fit shifts of  0\farcs11$\pm$0\farcs12 and 0\farcs07$\pm$0\farcs12 in right ascension and declination, respectively. Our computed coordinates of PSR\,  J0007+7303 are $\alpha =00^{\rm h}  07^{\rm m} 01\fs59$; $\delta  = +73^\circ 03\arcmin 08\farcs10$,  well consistent with those of Halpern et al.\ (2004), with an attached nominal absolute uncertainty of 0\farcs6 (at the 90\% confidence level\footnote{{\texttt http://cxc.harvard.edu/cal/ASPECT/celmon/}}).
We note that the \chan\ coordinates of \psr\ are relative to epoch 2003.28, i.e. about 6.5 and 8.5 years before our optical observations (see Table\ref{obs}),
and are affected by an intrinsic unknown uncertainty owing to the unknown pulsar proper motion.
No more recent \chan\ observations of \psr\ have been performed after that of Halpern et al.\ (2004).  A $\gamma$-ray timing position of \psr\ has been recently derived from the analysis of the first two years of {\em Fermi}/LAT data (Abdo et al.\ 2012) and is $\alpha =00^{\rm h}  07^{\rm m} 01\fs7$ ($\pm 0\fs2$); $\delta  = +73^\circ 03\arcmin 07\farcs4$ ($\pm0\farcs8$) at epoch 2009.328, i.e. closer to the epoch of our observations. The {\em Fermi}/LAT $\gamma$-ray timing  error circle overlaps the \chan\ one, with its centroid being slightly offset by $\sim 1\farcs2$ to the south east with respect to the latter. In the following, we conservatively explored both positions in our search for the \psr\ counterpart.

Fig.\ 1 (top) shows a section of the GTC image centred on the pulsar position. No object is detected within the estimated \chan\ and {\em Fermi}/LAT error circles. Smoothing the image with a Gaussian kernel does not yield evidence of detection above the background fluctuations.  A 
faint object, of magnitude $r'\approx 27$, is detected $\sim 3\arcsec$  north west of the \chan\ position.
However, its positional offset
clearly proves that it  cannot be associated with \psr. 
 Indeed, such an offset would imply,
for the pulsar distance of $1.4 \pm 0.3$ kpc (Pineault et al.\ 1993),  
a transverse velocity of $\sim 2400$ km s$^{-1}$, well beyond the far end of the pulsar velocity distribution (e.g., Hobbs et al.\ 2005). Moreover, for a characteristic age of $\sim13.9$ kyr,  such an offset would also imply that \psr\  was born $\approx 1.3^{\circ}$ south east of its current position, i.e. outside the CTA\,1  SNR. 
Thus, \psr\ is undetected in our images, and we computed the upper limit on its optical flux.  Following a standard approach (e.g., Newberry\ 1991), we determined the number of counts corresponding to a $3 \sigma$ detection limit  from the standard deviation of the background,  sampled around the \chan\  and {\em Fermi}/LAT error circles, and using a photometry  aperture of 0\farcs9 diameter ($\sim$7 pixels), equal to the image FWHM.   After applying the aperture correction, computed  from the PSF of a number of relatively bright but unsaturated field stars, and correcting for the airmass using the atmospheric extinction  coefficients for the Roque de Los Muchachos Observatory (Kidger et al.\ 2003), we then derived a $3 \sigma$  limit of $r' \sim 27.6$,  the deepest ever obtained for this pulsar.  \psr\ is obviously not detected in the less deep INT, TNG, and WHT $R$-band images and is also not detected in the TNG and WHT (Fig.1, bottom) $V$-band images. Similarly, we derived $3\sigma$ upper limits of $R\sim25.6$ (INT), $V\sim 26.1$, $R\sim26.2$ (TNG), and $V\sim 26.9$, $R\sim26.5$ (WHT).

\section{Discussion}

We compared our  $r'$ and $V$-band flux upper limits measured with the GTC and the WHT, respectively, and the published $B$-band one (Halpern et al.\ 2004), with the extrapolation in the optical domain of the model  which best fits the \xmm\ spectrum of \psr\ (Caraveo et al.\ 2010).  This is a double-component model consisting of a power-law (PL) with photon index $\Gamma_X= 1.30 \pm 0.18$ and a blackbody (BB) with temperature $kT=0.102^{+0.032}_{-0.018}$ keV, likely produced from a hot polar cap of radius $0.64^{+0.88}_{-0.20} d_{1.4}^2$, where $d_{1.4}$ is the pulsar distance in units of 1.4 kpc.  We corrected the optical flux upper limits for the interstellar reddening upon the hydrogen column density  derived from the best-fit: $N_{\rm H}=1.66^{+0.89}_{-0.76} ~10^{21}$cm$^{-2}$. Using the Predhel \& Schmitt (1995) relation and the interstellar extinction coefficients of Fitzpatrick (1999) 
we obtained extinction-corrected flux upper limits of 0.063 $\mu$Jy (r'), 0.144 $\mu$Jy (V), and 0.944 $\mu$Jy (B) for the best-fit value of the $N_{\rm H}$.
Fig.\ 2 shows the optical--to--X-ray spectral energy distribution of \psr. As seen, our deepest optical upper limit is consistent with the 90\% uncertainty of the extrapolation of the PL component, suggesting that both the optical and X-ray emission might be consistent with a single spectral model and be produced by the same mechanism at variance with, e.g. the Vela pulsar where a clear break is visible between the two spectral regions (Mignani et al.\ 2010c). 
According to the best-fit spectral parameters, the non-thermal unabsorbed X-ray flux of \psr\  in the 0.3--10 keV energy range is $F_{\rm X} = (0.686 \pm 0.100) \times 10^{-13}$ erg cm$^{-2}$ s$^{-1}$ (Caraveo et al.\ 2010). For the largest value of the inferred interstellar extinction ($A_r = 0.71^{+0.38}_{-0.33}$), the upper limit on the non-thermal unabsorbed $r'$-band flux (assuming that the \psr\ optical flux is dominated by  non-thermal emission, as observed in young pulsars) is  $F_{\rm opt} \sim 1.22 \times 10^{-16}$ erg cm$^{-2}$ s$^{-1}$. This gives an unabsorbed non-thermal optical--to--X-ray flux ratio $F_{\rm opt}/F_{\rm X} \la  1.56 \times 10^{-3}$, consistent with that inferred for other young  pulsars (e.g., Zharikov et al.\ 2006).

We also compared our optical upper limits with the extrapolation of the {\em Fermi}/LAT $\gamma$-ray spectrum of \psr. Since there is a marginal indication of a change in the pulsar spectral parameters after the 2009 May glitch, we assumed the post-glitch spectral parameters (see Tab.\ 3 of Abdo et al.\ 2012) that correspond to an epoch closest in time to  our optical observations. The $\gamma$-ray spectrum is best-fit by a PL with photon index $\Gamma_{\gamma}=1.50 \pm 0.03 \pm 0.04$ plus an exponential cut-off at an energy $E_c= 4.74 \pm 0.28 \pm 0.79$ GeV, where, in both cases, the first and second errors are statistical and systematical, respectively.  Our $r'$-band upper limit lies well below the extrapolation of the $\gamma$-ray PL confirming the presence of a break in the $\gamma$-ray spectrum, already made evident by the comparison with the X-ray PL.  Similar breaks are observed in many {\em Fermi} pulsars (e.g., Mignani et al.\ 2010a; 2011; 2012) but  not, e.g. in PSR\, J1048$-$5832 (Razzano et al.\ 2012). Finally, we note that the ratio between the unabsorbed optical flux and the $\gamma$-ray one for \psr\ is $F_{\rm opt}/F_{\gamma} \la  5.9 \times 10^{-6}$. This limit falls in the lowest 30\% bound among all {\em Fermi} pulsars that are either detected in the optical or for which significantly deep optical upper limits exist (The {\em Fermi}/LAT Collaboration, in prep.). This suggests that \psr\ might be one of the intrinsically faintest pulsars in the optical with respect to the $\gamma$-rays.

\begin{figure}
\centering
{\includegraphics[height=8cm, angle=270,clip=]{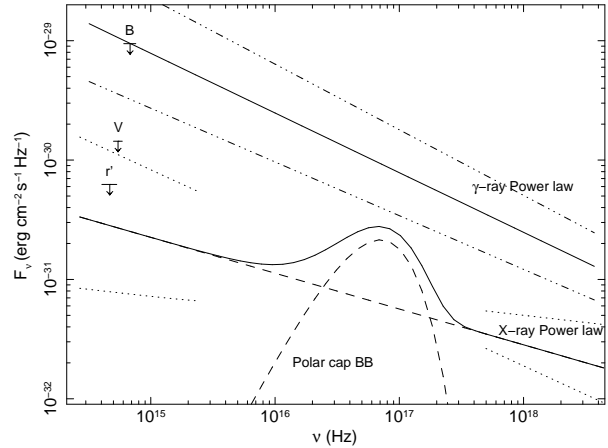}}
\caption{\label{sed} 
Extinction-corrected optical flux upper limits for \psr\ compared with  the extrapolation in the optical domain  of the PL+BB spectrum (solid line) that best-fits the {\em XMM-Newton} data (Caraveo et al.\ 2010). The two X-ray spectral components are shown by the dashed lines and labeled. The dotted lines correspond to the 90\% errors on the extrapolation of the X-ray PL component.  The plot also shows the extrapolation of the PL+exponential cutoff spectrum that best-fits the {\em Fermi}/LAT $\gamma$-ray data (Abdo et al.\ 2012) and the associated 90\% errors (dot-dashed lines). }
\end{figure}

We used our deepest upper limit in the $r'$ band to constrain the optical luminosity of \psr.  For the largest values of the distance ($1.4 \pm 0.3$ kpc; Pineault et al.\ 1993) and interstellar extinction, this limit corresponds  to an optical luminosity $L_{\rm opt} \la 4.24 \times 10^{28}$ erg cm$^{-2}$ s$^{-1}$. This implies an optical emission efficiency $\eta_{\rm opt} \equiv L_{\rm opt}/\dot{E} \la 9.4 \times 10^{-8}$.  This limit is well above the optical emission efficiency of the Vela Pulsar ($\eta_{\rm opt} \sim  2.9 \times 10^{-9}$) and below the limits for other Vela-like pulsars observed in the optical, such as PSR\, B1706$-$44, PSR\, J1357$-$6429, PSR\, J1028$-$5819	 (Mignani et al.\ 1999; 2011; 2012) and  PSR\, J1048$-$5832 (Razzano et al.\ 2012), which have been derived, however, from less deep optical observations. Thus, our result confirms that, in general, the optical emission efficiency of Vela-like pulsars is much lower than that of the Crab-like ones. However, we cannot yet determine whether the very low optical emission efficiency of the Vela pulsar is symptomatic of an even more dramatic change in the intrinsic pulsar optical emission output around  characteristic ages of 10--20 kyrs, or it is peculiar to this object only. 
Settling this issue obviously passes through the optical identification of at least some of these Vela-like pulsars. Unfortunately, although they have been all observed with the Very Large Telescope (VLT), for many of them the observations have been hampered by the presence of nearby bright stars. Thus, their identification can be better pursued either with the {\em HST} or with adaptive optics devices, assuming a spectral rise towards the near-infrared. 

\section*{Acknowledgments}
Based on observations made with the Gran Telescopio Canarias (GTC), installed in the Spanish Observatorio del Roque de los Muchachos of the Instituto de Astrof'sica de Canarias, in the island of La Palma. The research leading to these results has received funding from the European Commission Seventh Framework Programme (FP7/2007-2013) under grant agreement n. 267251. NR, DFT, DH, and AC research done in the framework of the grants AYA2009-07391,
AYA2012-39303, SGR2009- 811, and iLINK2011-0303. We thank the referee, Jules Halpern, for his useful comments that contributed to improve our manuscript.

\label{lastpage}


\begin{thebibliography}{99}

\bibitem[\protect\citeauthoryear{Abdo et al.}{2008}]{abdo08} Abdo A., Ackermann M., Atwood W. B., et al., 2008, Science, 322, 1218

\bibitem[\protect\citeauthoryear{Abdo et al.}{2012}]{abdo12} Abdo A., Wood K.S., DeCesar M. E., et al., 2012, ApJ, 744, 146

\bibitem[\protect\citeauthoryear{Aliu et al.}{2012}]{aliu12} Aliu E., Archambault S., Arlen T., et al., 2012, ApJ, in press, arXiv:1212.4739

\bibitem[\protect\citeauthoryear{Bertin \& Arnouts}{1996}]{bs96}Bertin E. \& Arnouts S., 1996, A\& A Suppl., 117, 393

\bibitem[\protect\citeauthoryear{Camilo et al.}{2009}]{cam09} Camilo F., Ray P. S., Ransom S. M., et al., 2009, ApJ, 705,1

\bibitem[\protect\citeauthoryear{Caraveo et al.}{2010}]{car10} Caraveo P.A., De Luca A., Marelli M., et al., 2010, ApJ, 725, L6

\bibitem[\protect\citeauthoryear{Collins et al.}{2011}]{col11}Collins S., Shearer A., Mignani R. P., 2001, in Radio pulsars: an astrophysical key to
unlock the secretes of the universe, AIP Conf. Proc., 1357, 310

\bibitem[\protect\citeauthoryear{Fitzpatrick}{1999}]{fit99} Fitzpatrick E. L. 1999, PASP, 111, 63

\bibitem[\protect\citeauthoryear{Halpern et al.}{2004}]{hal04} Halpern J. P., Gotthelf E.V., Camilo F., Helfand D. J., Ransom S. M., 2004, ApJ, 612, 398

\bibitem[\protect\citeauthoryear{Hobbs et al.}{2005}]{hob05}Hobbs G.,  Lorimer D. R., Lyne A. G., Kramer M., 2005, MNRAS, 360, 974

\bibitem[\protect\citeauthoryear{Kidger et al.}{2003}]{kid03} Kidger M. R.,  Fabiola M.-L., Narbutis D., Perez-Garcia  A., 2003, The Observatory, Vol. 123, p. 145-150

\bibitem[\protect\citeauthoryear{Lattanzi et al.}{1997}]{lat97}Lattanzi M.~G., Capetti A., \& Macchetto F.~D.\ 1997, A\&A, 318, 997 

\bibitem[\protect\citeauthoryear{Mignani et al.}{1999}]{mig99} Mignani R.P., Caraveo P.A., Bignami G.F.,  1999, A\&A, 343, L5

\bibitem[\protect\citeauthoryear{Mignani et al.}{2010a}]{mig10a} Mignani R.P., Pavlov G.G., Kargaltsev O., 2010a, ApJ, 720, 1635

\bibitem[\protect\citeauthoryear{Mignani et al.}{2010b}]{mig10b} Mignani R. P., Jackson A. C., Spiers A, 2010b, A\&A, 520, 21

\bibitem[\protect\citeauthoryear{Mignani et al.}{2010c}]{mig10c} Mignani, R.P.,  Sartori, A., De Luca, A., et al., 2010c, A\&A, 515, 110

\bibitem[\protect\citeauthoryear{Mignani et al.}{2011}]{mig11} Mignani R.P., Shearer A., De Luca A.,  Moran P., Collins S., Marelli M., 2011, A\&A, 533, 101

\bibitem[\protect\citeauthoryear{Mignani et al.}{2012}]{mig12} Mignani R.P., Razzano M., Esposito P., De Luca A., Marelli M., Oates S. R.,  Saz-Parkinson P., 2012, A\&A, 543, 130

\bibitem[\protect\citeauthoryear{Monet et al.}{1998}]{mon98} Monet D., et al., 1998, A Catalogue of Astrometric Standards: USNO-SA2.0 (Flagstaff: US Naval Obs.)

\bibitem[\protect\citeauthoryear{Newberry}{1991}]{new91} Newberry M.V., 1991, PASP, 103, 122

\bibitem[\protect\citeauthoryear{Nolan et al.}{2012}]{nol12} 	Nolan P.L., Abdo A. A., Ackermann M., et al., 2012, ApJS, 199, 31

\bibitem[\protect\citeauthoryear{Pineault et al.}{1993}]{pin93} Pineault S., Landecker T. L., Madore B.,  Gaumont-Guay S., 1993, AJ, 105, 1060
	
\bibitem[\protect\citeauthoryear{Predehl, P. \& Schmitt}{1995}]{ps95} Predehl P. \& Schmitt J.H.M.M. 1995, A\&A, 293, 889

\bibitem[\protect\citeauthoryear{Razzano et al.}{2012}]{raz12}Razzano M., Mignani R.P., Marelli M., De Luca, A., 2012, MNRAS, in press, arXiv:1210.7730

\bibitem[\protect\citeauthoryear{Seward et al.}{1995}]{sew95} Seward F.D., Schmidt B., Slane P., 1995, ApJ, 453, 284

\bibitem[\protect\citeauthoryear{Shearer et al.}{2012}]{she12} Shearer A., Mignani R. P., Collins S., et al., 2012, MNRAS, to be submitted 

\bibitem[\protect\citeauthoryear{Skrutskie et al.}{2006}]{skr06}  Skrutskie M.~F., Cutri R.~M., Stiening R., et al.\ 2006, AJ, 131, 1163 

\bibitem[\protect\citeauthoryear{Slane et al.}{1997}]{sla97} Slane P., Seward F. D., Bandiera R., Torii K.,  Tsunemi H., 1997, ApJ, 485, 221

\bibitem[\protect\citeauthoryear{Slane et al.}{2004}]{sla04} Slane P., Zimmerman E. R., Hughes J. P., Seward F. D., Gaensler B. M., Clarke M. J., 2004, ApJ, 601,1045

\bibitem[\protect\citeauthoryear{Zharikov et al.}{2006}]{zha06} Zharikov S. V., Shibanov Yu. A., \& Komarova V. N., 2006, AdSpR, 37, 1979

\end{thebibliography}
\end{document}